\documentclass[trackchanges, twocolumn]{aastex701}

\usepackage{amsmath}

\accepted{March 18, 2026}
\submitjournal{The Astrophysical Journal}

\begin{document}

\title{Global Coronal Equilibria with Solar Wind Outflow II -- Optimizing the Outflow Model}

\author[orcid=0000-0002-5217-6361,sname='Rice']{Oliver E. K. Rice}
\affiliation{Department of Mathematical Sciences, Durham University, Durham, DH1 3LE, UK}
\email[show]{oliver.e.rice@durham.ac.uk}  

\author[orcid=0000-0002-2728-4053,gname=Bosque, sname='Yeates']{Anthony R. Yeates} 
\affiliation{Department of Mathematical Sciences, Durham University, Durham, DH1 3LE, UK}
\email{anthony.yeates@durham.ac.uk}

\begin{abstract}

We expand upon our paper \cite{2021ApJ...923...57R} which introduced `Outflow Fields': alternatives to the widely-used potential field source surface (PFSS) extrapolations of the Sun's coronal magnetic field which take into account the effect of the solar Wind. We showed that our fields have several advantages over PFSS, namely more accurate measurements of the Open Solar Flux (OSF) relative to observations, more realistic streamer shapes and less dependence on the arbitrary source-surface height. In this paper we seek to quantify these improvements. This includes comparison of magnetic field line angles with eclipse photography, an improved solar wind solution model and the introduction of data from a wider range of observations. We use these comparisons to determine the optimum parameters for our model using an evolutionary algorithm, in addition to the creation of synthetic eclipse images. We find that our Outflow Fields can accurately capture the overall topology of the magnetic field, and reduce the well-known discrepancy with in-situ magnetic field measurements by a significant margin relative to PFSS. Specifically, over the period between $2000$ and $2022$ for a typical source-surface height we find that optimized Outflow fields reduce this discrepancy from around $45\%$ to $24\%$ while also matching the field line topology seen during eclipse photography. Our model is presented for wider use by the community as a new python package `outflowpy'. 

\end{abstract}



\section{Introduction}\label{sec:introduction}

Numerical models of the solar magnetic field exist in a vast array of guises, varying from 2.5D models of individual features on the solar surface, through to hugely complex thermodynamically-accurate representations of the entire corona  \citep[see reviews by ][]{2012LRSP....9....6M, 2017SSRv..210..249W, 2018LRSP...15....4G}. Historically, one of the more popular has been the PFSS (potential field source surface) model \citep{1969SoPh....9..131A, 1969SoPh....6..442S}, which is perhaps the simplest of all global coronal models. This assumes a current-free coronal magnetic field which becomes purely radial at a specified height, known as the source surface. These models have long been used as a cheaply-calculated first approximation, either in their own right or as initial conditions for more complex simulations. Of particular note is that PFSS models (with suitable modifications) are widely used as a basis for space weather predictions \citep{2018SpWea..16.1644M}, especially in combination with the Schatten Current Sheet model \citep{1971CosEl...2..232S} or in the form of the WSA model, often used for determining solar wind speeds \citep{2017HGSS....8...21S}. 

In our previous paper \citep[][hereafter Paper I]{2021ApJ...923...57R} we introduced a variation on these PFSS fields, named `Outflow Fields', which are similarly fast to compute but have numerous advantages. The most significant of these we claim are as follows:
\begin{enumerate}
    \item The well-known discrepancy (the `Open Flux Problem') between the overall Open Solar Flux (OSF) predicted by PFSS models and in-situ observations \citep{2017ApJ...848...70L} is reduced when using an Outflow Field. The OSF is defined as the total unsigned radial magnetic flux passing through a given altitude (for our purposes, the source-surface height).
    \item When the source surface height is sufficiently high, the structure of an Outflow Field is roughly independent of this height (which is usually somewhat arbitrarily chosen), as instead the magnetic field structure is determined primarily by a physically-motivated solar wind outflow profile.
    \item The shapes of streamers and pseudo-streamers in the lower corona match those observed during solar eclipses much more closely compared to PFSS fields.    
\end{enumerate}
In this paper we seek to further justify and quantify these claims.

A detailed discussion of the motivation behind and the calculation of the Outflow Fields is contained within Paper I -- almost all numerical schemes remain unchanged in this new work. In brief, the Outflow model seeks to find equilibria of the magneto-frictional (MF) model  \citep[e.g.][]{1986ApJ...309..383Y,2006ApJ...641..577M, 2022GApFD.116..305Y}, including a term representing the solar wind, but without the need for expensive time-stepping. In this way the currents in the upper corona caused by the solar wind are accounted for, whereas those lower down (which lead to structures such as flux ropes) are ignored. Mathematically, where a PFSS field solution $\bf{B}$ can be expressed as a vector potential:
\begin{equation}
    {\bf B} = \nabla \Phi,
\end{equation}
for some scalar field $\Phi$, instead an Outflow Field allows more radial variation:
\begin{equation}
\textbf{B} = f(r) \nabla \Phi,
\end{equation}
with $f(r)$ and $\Phi$ determined such that $\bf{B}$ satisfies the magneto-frictional equilibrium equation 
\begin{equation}
\left(\frac{(\nabla\times{\bf B})\times{\bf B}}{\nu_0 \lvert{\bf B}\lvert^2} + v(r) \mathbf{e}_r\right)\times{\bf B} ={\bf 0},
\label{eqn:mfequ}
\end{equation}
where  $\nu_0$ is the magnetofrictional constant, and $\mathbf{e}_r$ is the unit vector in the radial direction. In Paper I we outlined in detail a numerical method to achieve this, while preserving the divergence-free condition precisely. The constant $\nu_0$ and the solar wind speed profile $v(r)$ are essentially free physically-motivated parameters. In this paper (where necessary) we use distance units of solar radii ($R_\odot$) and time units of seconds ($s$), but for most purposes $v(r)$ and $\nu_0$ only ever appear multiplied together, and can be combined with a distance factor to create a dimensionless `speed' $V(r) = R_\odot \nu_0 v(r)$. This measure of the relative strength of the outflow compared to the frictional relaxation rate is used as a proxy for the solar wind speed throughout the paper.

The optimal radial profile and overall magnitude of $V(r)$ are the topic of much discussion in this paper, as they alter the nature of the solutions considerably. But it is important to note that only the product $V(r)$ can be empirically determined by fitting the model to observations. Thus the model cannot be used to predict a dimensional solar wind speed $v(r)$ unless an appropriate value of $\nu_0$ can be independently determined. We initially choose $\nu_0 = 5 \times 10 ^ { -17} \rm s \, cm ^{-2}$ because this value has been used in recent magnetofrictional models \citep[e.g.][]{2024SoPh..299...83Y}, although we find later in this paper that a lower value of $\nu_0$ may give more realistic speeds when $V(r)$ is optimized against observations.

Our intention is for the Outflow model to add to the existing suite of available global coronal models, by filling a niche between the simplicity of PFSS and the complexity of either (i) time-dependent models -- such as magnetofriction \citep[MF;][]{2010JGRA..115.9112Y} or magneto-hydro-dynamics \cite[MHD;][]{2018JSWSC...8A..35P, 2021A&A...653A..92P, 2022A&A...662A..50V, 2023ApJ...959...77L, 2025A&A...694A.306B} -- or (ii) static extrapolations which require more complex boundary data or additional assumptions, like magnetohydrostatic or force-free extrapolations \citep{2020SoPh..295..145W, 2021LRSP...18....1W}. 

The main advantage of our model is that it requires no more input data than PFSS (apart from an assumed solar wind speed), but does provide more accurate results in several respects, as stated above. Naturally there exist more complex models which allow for better representations of physical reality, but generally these require both more boundary data (which is often very difficult to obtain compared to the radial magnetic fields we require) and take far longer to compute. We do not propose that Outflow fields are an alternative to such simulations, as they cannot possibly represent dynamic or small-scale magnetic field behavior, but do assert that they may (in many scenarios) be used like-for-like in place of PFSS. 

Indeed, for example, \cite{2025arXiv251119975L} use Outflow models to initialize MHD simulations in place of the standard PFSS initial condition, and find that their MHD solutions converge to an equilibrium solution far more quickly than they otherwise would. Outflow fields were also compared like-for-like with PFSS in \cite{2025ApJ...985..190W}, who use our model (named in that paper as `SEMF') to investigate the effects of the solar wind on the open-closed magnetic field boundary, albeit with relatively modest solar wind speeds. 

We believe that our work is the first which seeks to find equilibrium solutions of the form of Equation \eqref{eqn:mfequ}, but we are not the first to attempt to modify PFSS to improve comparisons with observations. Of particular note is the work of \cite{2023Physi...5..161T}, who also present a non-potential steady state magnetic field solution which qualitatively looks very similar to the Outflow model, and also shows an increase in OSF relative to PFSS. They modify the established technique of \cite{1986ApJ...306..271B}, which artificially introduces a non-zero current density in the plane perpendicular to the radial direction. This is essentially equivalent to finding solutions of Equation \eqref{eqn:mfequ} but `uncrossed' with $\textbf{B}$, allowing for more freedom in the solutions but also being less physically-motivated by an explicit solar wind speed. 

We begin in Section \ref{sec:windspeed} by discussing in detail the isothermal Parker solar wind speed solutions, and how they can be incorporated into Outflow models. Unlike in Paper I, where we used an explicit approximation for the solar wind profile, our new approach allows for higher solar wind speeds by allowing the profile to be non-zero near the solar surface. Relaxing this assumption allows for more variation in predicted streamer shapes, allowing us to fit the model more closely to observations.

In Section \ref{sec:heightvary} we examine in detail our claim that Outflow Fields have less dependence on the source-surface height $r_{ss}$ than PFSS when regarding the total Open Solar Flux (OSF). We also outline the steps used to prepare data from the HMI and MDI instruments for use with our models. We further compare these predictions to in-situ measurements of the OSF using the data from the OMNI spacecraft, corrected for various effects by \cite{2022SoPh..297...82F}, examining the overall shortfall in open flux predictions and the correlations between the two data sets.

In the remaining part of the paper we take a different approach, seeking to optimize the free parameters $\nu_0$ and $v(r)$ such that the overall topology of the predicted magnetic field closely approximates photographs of the lower corona taken during solar eclipses (we consider 12 eclipses in the period 2006-2024). Automatic detection of the topology of the magnetic field during eclipses has been achieved by \cite{2020ApJ...895..123B} by using a Rolling Hough Transform, and used in combination with PFSS models by \cite{2024AGUFMSH51D2931B}. A similar approach by \cite{2025ApJ...995...75U}, called `QRaFT', seeks to achieve the same goal of detecting the overall topology of the magnetic field in both real and synthetic eclipse photographs. With goals similar to our own, \cite{2025arXiv250313292R} used QRaFT in combination with statistical measures with the ultimate goal of optimizing MHD models. Alternatively, \cite{2023ApJ...943..124P} use observations of the 2019 eclipse by manually identifying the overall streamer shapes to determine the best source data for space weather predictions.  

Compared to these works we use a different approach to detect the direction of the magnetic field in the plane of view, but a similar analysis of the field line angles. We combine this analysis with an evolutionary algorithm (CMA-ES) to determine the ideal parameters for our Outflow model, seeking to match the average field line angles at all altitudes up to $2.5R_\odot$. We find that these optimized Outflow fields outperform PFSS equivalents both in overall magnetic field topology and predictions of OSF. 

We present the code used to generate Outflow Fields (along with our edge detection and synthetic image generation algorithms) as an open-source python package `outflowpy' \citep{outflowpy}, available at \url{https://github.com/oekrice/outflowpy}, which is designed to be easy to use and as compatible as possible with the existing `pfsspy' \citep{Stansby2020} package (used for calculating PFSS fields). 

\section{Isothermal Parker Solar Wind Speeds}
\label{sec:windspeed}

In Paper I, we exclusively used an explicit approximation to the isothermal Parker solar wind solution \citep{1958ApJ...128..664P}. The exact solution can be expressed implicitly as
\begin{equation}
    \left(\frac{v(r)}{v_c}\right)^2 - 2\ln\left(\frac{v(r)}{v_c}\right) = 4\ln\left(\frac{r}{r_c}\right) + \frac{4r_c}{r} - 3. \label{eqn:parker_implicit}
\end{equation}
In that paper we assumed the isothermal sound speed $c_s$ was low enough that the critical radius $r_c$ was sufficiently high relative to the computational domain to use the explicit approximation
\begin{equation}
v_{\rm out}(r) = v_1\frac{r_1^2\mathrm{e}^{-2r_c/r}}{r^2\mathrm{e}^{-2r_c/r_1}}, \label{eqn:parker_approx}
\end{equation}
where $r_1$ and $v_1$ are the radius of the upper boundary, and the velocity at that height, respectively.

Here, for clarity, we calculate the sound speed $c_s$ and the critical radius $r_c$ with the formulae:
\begin{align}
    c_s &= \sqrt{\frac{k_bT_0}{m_p}} \approx \sqrt{\frac{1.38 \times 10^{-23} \cdot T_0}{1.67 \times 10 ^{-27}}} {\rm m}{\rm s}^{-1}, \label{eqn:soundspeed} \\
    r_c &= \frac{GM_\odot}{2c_s^2} \approx \frac{6.67 \times 10^{-11} \cdot 1.19 \times 10^{30}}{2c_s^2} \rm m,
\end{align}
where $k_b$ is the Boltzmann Constant, $m_p$ is the proton mass, $G$ is the gravitational constant, $M_\odot$ is the mass of the Sun and $T_0$ is the isothermal coronal temperature in Kelvin. 

It is important to note that despite our use of the Parker model as a first approximation to a realistic outflow profile, in reality the nature of the solar wind (and in particular its dependence on the temperature) is far more complex. Of particular note, it appears from in-situ observations that the solar wind is faster when originating from cooler regions of the corona, in direct contradiction to Equation \eqref{eqn:soundspeed} \citep[e.g.][]{2003JGRA..108.1158G, 2003Sci...302.1165S, 2021ApJ...911L...4H}. Alas, our model currently requires a solar wind profile which depends only on altitude, so these effects cannot be easily or realistically incorporated.

Later in this paper we will remove the assumption of a Parker wind profile entirely, but for the initial illustrations of the effects of adding an outflow to the PFSS model we retain the overall `shape' of the Parker solution, but without reference to the coronal temperature.

For a coronal sound speed of $100\,\mathrm{km}\,\mathrm{s}^{-1}$ (which is on the low end of such estimates), the critical radius is around $5.7R_\odot$, which is high enough for the approximation (\ref{eqn:parker_approx}) to be valid (as assumed in Paper I). However, for higher sound speeds the critical radius decreases such that \eqref{eqn:parker_approx} ceases to be reasonable even at altitudes below $2.5 R_{\odot}$. Thus in this paper we instead solve the implicit Equation \eqref{eqn:parker_implicit} exactly. Reliably selecting the `correct' fast wind solution is not trivial, and we have written a bespoke solver to ensure that this is done reliably for reasonable input parameters. A similar approach was used in \cite{2025arXiv251119975L}, wherein Outflow Fields were used to initialize MHD simulations, but the solar wind speed was solved explicitly using Lambert Functions instead.

\begin{figure*}
\plotone{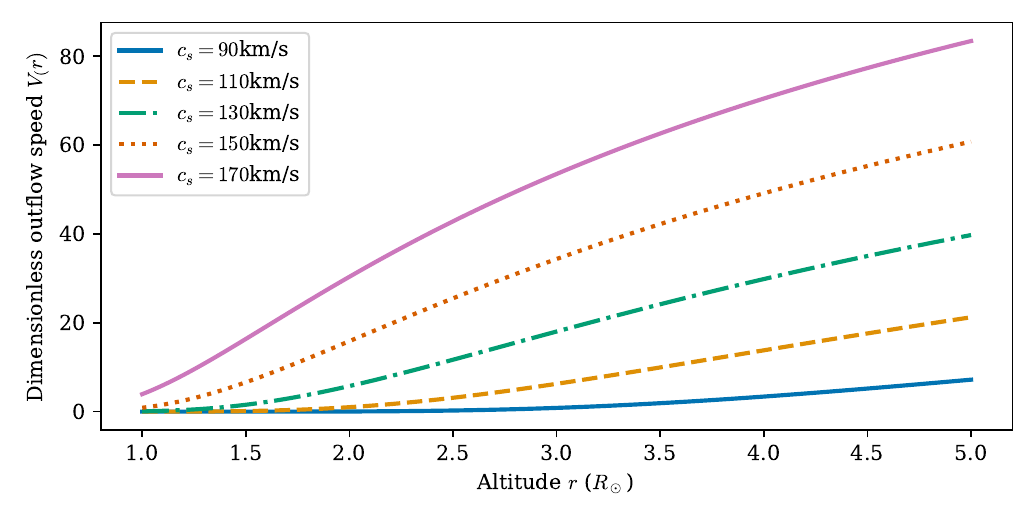}
\caption{Outflow speeds $V(r)$ as a function of altitude $r$, shown up to five solar radii and for a variety of coronal sound speeds $c_s$, calculated using the implicit formula \eqref{eqn:parker_implicit} and a magnetofrictional constant $\nu_0 = 5 \times 10 ^ { -17} \rm s \, cm ^{-2}$.}
\label{fig:1_outflow_speeds}
\end{figure*}

Figure \ref{fig:1_outflow_speeds} shows the outflow speed for various isothermal sound speeds up to five solar radii. This speed is expressed as the dimensionless product $V(r) = R_\odot\nu_0 v(r)$.
 
In this plot (and as in our original paper) we choose a magnetofrictional constant of $\nu_0 = 5 \times 10 ^ { -17} \rm s \, cm ^{-2}$ \citep{2016A&A...594A..98Y}, but later in the paper we will determine the ideal values of $V(r)$ empirically without the need to assume a value for $\nu_0$. Essentially, the sound speed $c_s$ determines the overall shape of the solution, and the magnetofrictional constant $\nu_0$ scales this solution by a constant factor. Thus with this simple isothermal model we have two degrees of freedom with which to determine the outflow speed profile.

Note that for high sound speeds our original assumption that the outflow speed is negligible at $r=R_\odot$ is clearly no longer valid. This has implications for the numerical scheme described in Paper I, namely the lower boundary condition for the radial eigenfunctions $H(\rho)$, which is Equation (27) in that paper. To allow for $v(\rho_0) > 0$ we modify this boundary condition to become
\begin{equation}
    (H_l(\rho)e^\rho)^{'} \vert _{\rho_0} - v(\rho) e^{2\rho} H_l(\rho) = e^\rho,
\end{equation}
where $\rho$ is the stretched radial coordinate ($r = e^\rho$) and $\rho_0 = 0$ is the lower boundary. Note that in the unmodified condition the second term was neglected entirely. This modification allows for mathematically consistent solutions when the sound speed is high, and is the only difference in the numerical scheme between the calculations in this paper and those in Paper I.

\section{Dependence of Open Flux on the Source Surface Height}
\label{sec:heightvary}

We previously claimed that one of the major advantages of Outflow Fields relative to PFSS equivalents is that the open flux (OSF) through the source surface is less dependent on the altitude of the source surface -- this is a major issue with PFSS models. Most commonly the PFSS source surface height is chosen to be $r_{ss} = 2.5 R_\odot$.  However, there are many studies which regard this dependency as a feature, not a bug, and study the effects of changing this height (or indeed experimenting with a non-spherical source surface) with the goal of more closely matching observations of the coronal magnetic field structure \citep[eg.][]{1969SoPh....6..442S, 1982SoPh...79..203L, 2011SoPh..269..367L,  2014JGRA..119.1476A,2020ApJ...889L..28V,2025ApJ...993..242S,2025ApJ...988..239M}. 

These works generally agree that in order to match OSF measurements the source surface needs to be lower than $2.5 R_\odot$, although the ideal height varies depending on a number of factors -- in particular the stage of the solar cycle. The unpredictability of these variations renders such approaches unsuitable for predictive uses of PFSS. Indeed, whereas \cite{2019SpWea..17.1293N} suggests that the optimal source-surface height is always below $2.5R_\odot$ (and even as low as $1.25 R_\odot$ at solar minimum!) to match OSF and coronal hole measurements, \cite{2024ApJ...965L...1H} propose that an optimal height exceeds $2.5R_\odot$ at solar maximum, when comparing with MHD simulations. Moreover, in comparison with eclipse images, \cite{2024AGUFMSH51D2931B} deduced the opposite correlation, asserting that the optimum height varies from $1.3R_\odot$ at maximum to $3.0R_\odot$ at minimum.

In our Outflow model, by contrast, the shapes of the field lines are determined predominantly by the presence of the solar wind rather than the height of the source surface. As a result the magnetic field lines can become essentially radial at a height well below the source surface. Above this point the total OSF will be roughly constant, as none of the magnetic field lines loop back towards the solar surface. This effect can be observed in Figure 3 of Paper I.

To quantify this, in this Section we compare the OSF as a function of height, using PFSS and Outflow Fields calculated based on lower boundary magnetic field data obtained from the HMI synoptic maps \citep{2011SoPh..269..367L, 2018arXiv180104265S} around the time of the 2017 August 21 Solar Eclipse. Figure \ref{fig:3_plot_surface} shows this input magnetic field data. The upper panel shows the `raw' HMI data, which on this date is a composite of the synoptic maps of Carrington Rotations 2193 and 2194, aligned such that the center of the plot is the data measured on August 21 with each longitude weighted as a proportion of the two maps depending on the observation time (i.e. each longitude is averaged from two synoptic maps based on the proximity of its observation time to the center of each map). This is a compromise between using the synoptic maps in their raw state, which leads to a discontinuity on the far side of the Sun, and more complex methods using either full-disk magnetograms \citep[e.g.][]{2025ApJ...992...89S} or advanced filtering techniques \citep[e.g.][]{2015SoPh..290.1105H} -- or indeed a combination of both -- to predict the evolution of magnetic structures while they cannot be observed.

Using our approach the lower boundary condition has no discontinuities and can in a basic sense account for the flux emergence and decay between each synoptic observation. Including this time averaging step does not greatly alter the overall findings of our study -- there is less than a $1\%$ difference between the overall average OSF measurements when using the averaging as opposed to using the raw synoptic maps. We note, however, that including the effects of flux emergence and decay does reduce the daily fluctuations in OSF considerably, although by taking a 27-day average there is little effect overall and this does not have a major influence on our conclusions.

The raw data are too noisy and high in resolution to be of use as a lower boundary condition in themselves, so are processed as follows:

\begin{figure*}
\plotone{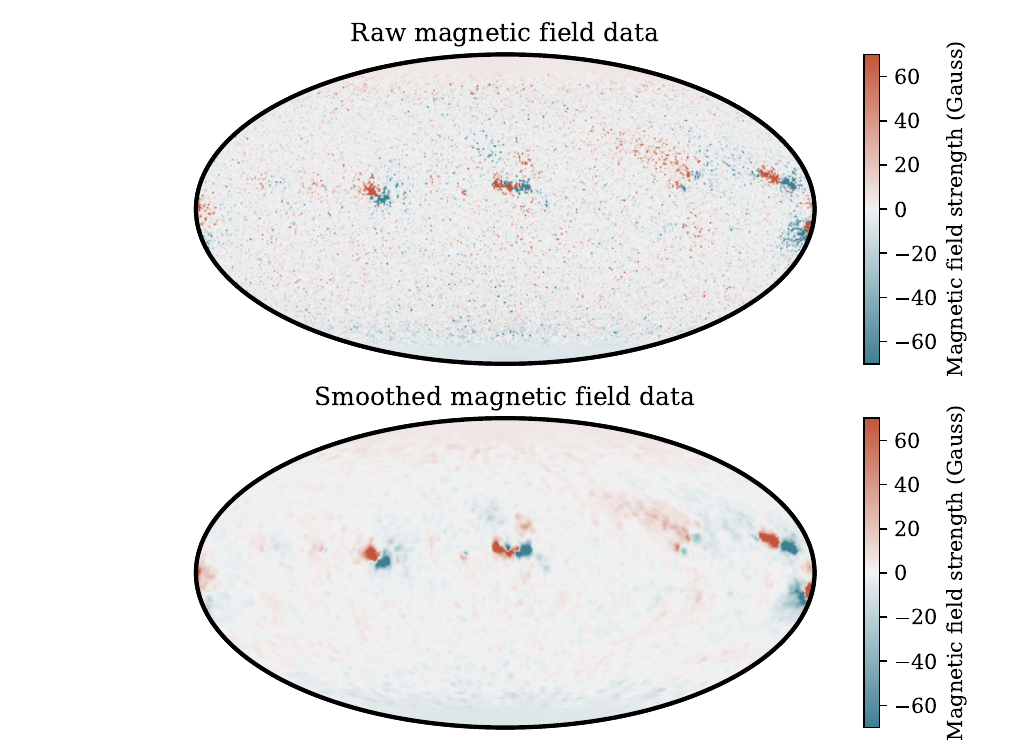}
\caption{Magnetic field data from the HMI synoptic map series, for the 2017 August 21. The upper panel shows the raw input data (a weighted combination of the synoptic maps for CR2193 and CR2194), and the lower panel shows this same data after our smoothing and interpolation process, ready to be used as a lower boundary condition for an Outflow Field.}
\label{fig:3_plot_surface}
\end{figure*}

\begin{itemize}
  \item Downsample the input data from a resolution of $3600 \times 1440$ to $1800 \times 720$ by simple averaging. This has very little effect other than speeding up the processing considerably.
  \item Apply a smoothing filter using a discrete Legendre Polynomial decomposition (as in the main outflow calculation). In the notation of Paper I, each eigenfunction coefficient $C_{l,m}$ is multiplied by a smoothing factor $e^{-\sigma \lambda_l}$, where $\lambda_l$ is the discrete eigenvalue of this mode (a numerical equivalent to $l(l+1)$) and $\sigma$ is a constant smoothing factor which we have empirically chosen to be $\sigma = 0.008 \Delta \phi$, where $\Delta \phi$ is the target grid resolution in the longitudinal direction. This essentially simulates applying the heat equation to the lower boundary data. Simpler methods, such as Gaussian smoothing or using an analytic Legendre decomposition, can lead to undesirable artifacts near the poles.
  \item Interpolate these data onto the required grid resolution (this allows for arbitrary grid shapes and sizes). The code uses a cubic bivariate spline approximation for this step. 
  \item Ensure the data are flux-balanced such that the overall magnetic flux through the lower boundary is precisely zero. To do this, we simply linearly scale all cells with positive flux such that the overall sum of the positive flux equals half the absolute sum of both polarities (and vice versa for negative flux). 
\end{itemize}

The equivalent magnetic field after applying this process is shown in the lower panel of Figure \ref{fig:3_plot_surface}.

\begin{figure*}
\plotone{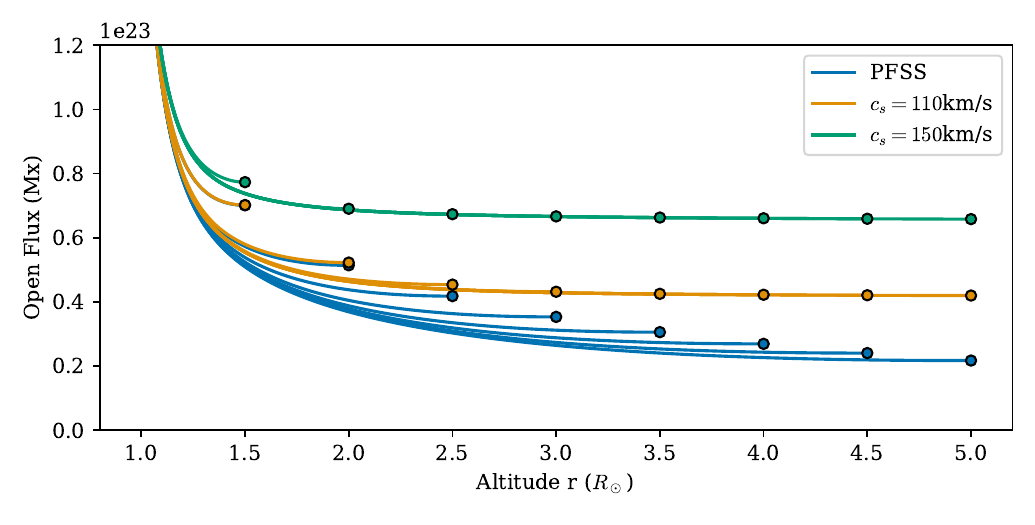}
\caption{Plots of the OSF on 2017 August 21 as a function of altitude, for both the PFSS model and Outflow Fields at both $110\,\mathrm{km}\,\mathrm{s}^{-1}$ and $150\,\mathrm{km}\,\mathrm{s}^{-1}$, with source surface heights $r_{ss}$ varying between $1.5 R_\odot$ and $5.0 R_\odot$. The source surface height for each result is indicated by the black-outlined circles at the end of each plotted line. The OSF is measured in Maxwells ($\mathrm{Mx}\equiv\mathrm{G}\,\mathrm{cm}^{-2}$), and the magnetofrictional constant used to generate the Outflow Fields is $\nu_0 = 5 \times 10 ^ { -17} \rm s \, cm ^{-2}$.}
\label{fig:2_openflux_comparison}
\end{figure*}

Figure \ref{fig:2_openflux_comparison} shows the OSF (total unsigned magnetic flux through $r=r_{ss}$) as a function of altitude for a selection of PFSS and Outflow Fields, using the magnetic field data shown in Figure \ref{fig:3_plot_surface} as a lower boundary condition. We compare the PFSS reference in blue with two series' of Outflow Fields at both $110\,\mathrm{km}\,\mathrm{s}^{-1}$ and $150\,\mathrm{km}\,\mathrm{s}^{-1}$ (both with $\nu_0 = 5 \times 10 ^ { -17}\,\mathrm{s}\,\mathrm{cm}^{-2}$ as the magnetofrictional constant). 

In each of these fields the OSF monotonically decreases with altitude as some field lines loop back  to connect to the solar surface. The OSF remains roughly constant above a certain radius for the Outflow Fields, which indicates that above this altitude the magnetic field is almost completely open. For each series of simulations, this graph shows the OSF for eight different source surface heights ($r_{ss}$), varying from $1.5 R_\odot$ to $5.0 R_\odot$. The black-outlined circles show the OSF through each of these source surfaces in turn.

We see that for the PFSS fields the OSF through the source surface decreases considerably and consistently as $r_{ss}$ increases. This is a well-known result \citep[noted originally upon the introduction of PFSS fields in ][]{1969SoPh....6..442S} and comes about because in PFSS models the streamers tend to extend upwards almost as far as the source surface height, and the topology of the overall magnetic field changes significantly as a result of changes to $r_{ss}$. In our Outflow Fields this is not the case. Instead, the presence of the solar wind is the main influence behind the magnetic field topology, and thus the OSF through $r_{ss}$ does not decrease significantly beyond a source-surface height of around $2.5 R_\odot$. This can be seen in the figure by the fact that the OSF profiles of the Outflow Fields are essentially identical with varying $r_{ss}$ height whereas all of the PFSS fields are individually quite clearly different. 

\begin{figure*}
\plotone{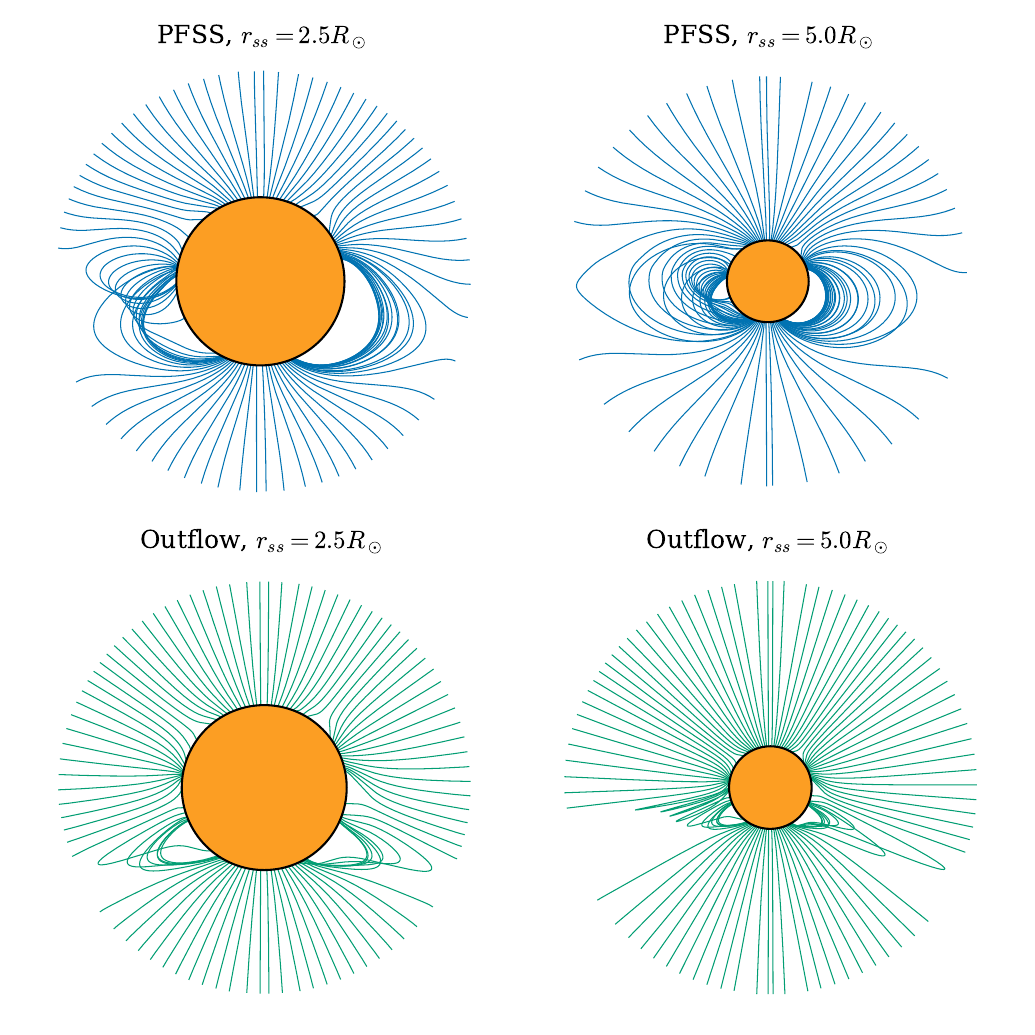}
\caption{Qualitative comparison of the magnetic field structures of PFSS fields and Outflow Fields (with $c_s = 150 \,\mathrm{km}\,\mathrm{s}^{-1}$) for two source-surface heights. The input magnetic field data is centered on the HMI measurements from 2017 August 21. The magnetofrictional constant used to generate the Outflow Fields is $\nu_0 = 5 \times 10 ^ { -17} \rm s \, cm ^{-2}$.}
\label{fig:4_plot_flines}
\end{figure*}

This can be seen qualitatively in Figure \ref{fig:4_plot_flines}, which shows the equivalent models with source surfaces at heights $r_{ss} = 2.5R_\odot$ and $r_{ss} = 5.0R_\odot$. The structures of the two PFSS fields are fundamentally different (even low down) as the closed-field regions expand significantly. The Outflow Fields are far more similar (at least up to $r_{ss} = 2.5R_\odot$), exemplifying the consistency of the model with varying source surface heights.

One major implication of this result is that in the Outflow model it is impossible to match OSF measurements (as in the works cited above) by altering the source surface height alone. Instead, using Outflow Fields we can change the outflow speed profile to accomplish this. This allows the magnetic field to be modeled accurately up to very high $r_{ss}$ without severe implications for the magnetic field topology low in the corona or the need to couple to different models for higher altitudes \citep[e.g. the Schatten current-sheet model,][]{2024FrASS..1176498K}.

We also see clearly from Figure \ref{fig:2_openflux_comparison} {that a higher outflow speed corresponds to a general increase in OSF: a result discussed extensively in our previous paper.

\section{Extrapolation comparisons with in-situ OSF}
\label{sec:frostmatch}

There is a well-known discrepancy between in-situ measurements of the total unsigned magnetic flux in interplanetary space (the `OSF') measured with instruments on spacecraft close to 1AU, and predictions of this flux based on magnetic field extrapolations from observations of the solar surface magnetic field. This is commonly known as the `open flux problem' \citep{2017ApJ...848...70L}. The established PFSS model underestimates the `true' OSF values often by up to a factor of two. There are many proposed suggestions as to why this is the case \citep[e.g.][]{2006JGRA..11110104O, 2022SoPh..297...82F, 2024ApJ...964..115A, 2025ApJ...987...32K}, but for the purposes of this section we will merely compare the performance of the Outflow model relative to PFSS.

As a `ground truth' for the total amount of open magnetic flux we use in-situ data from \citet{2022SoPh..297...82F}. This is collected from instruments aboard the \textit{Wind} and ACE spacecraft and provides perhaps the best estimate for the true topology of the heliospheric magnetic field near to 1AU. Notably, this dataset accounts for the influence of `switchbacks' \citep{2019Natur.576..228K, 2021A&A...650A...2D} whereby the magnetic field briefly kinks back upon itself. Switchbacks do not have any significant influence over the overall structure of the magnetic field, and will not be accounted for in either PFSS or Outflow extrapolations, and so being able to discount their contribution to the in-situ magnetic flux greatly improves the reliability of this comparison.

We make two key assumptions when comparing models extending to only a few solar radii with measurements near 1AU (approximately $215R_\odot$). The first is that the total magnetic flux remains constant throughout the region between the upper boundaries of our models and the Earth -- essentially equivalent to assuming that no magnetic field lines loop around and connect back to the Sun. The second is that the magnetic field strength at 1AU is latitudinally invariant \citep{1996GeoRL..23.3267S}. This is necessary as all in-situ measurements we use are at latitudes close to the Solar equator.

\begin{figure*}
\plotone{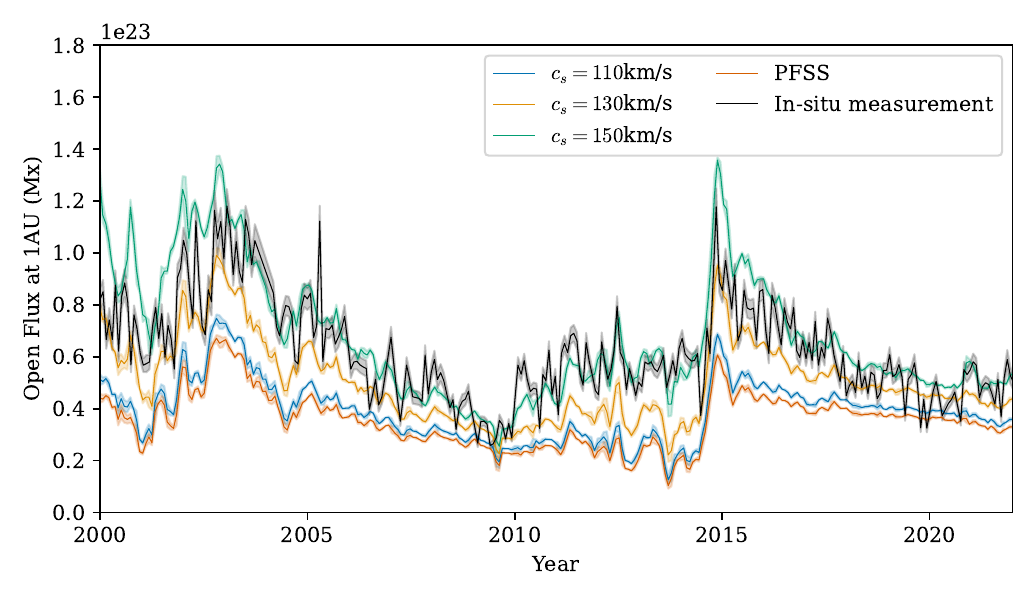}
\caption{Comparison between PFSS, Outflow Field extrapolations and in-situ measurements of the OSF at 1AU. The in-situ data is taken from \citet{2022SoPh..297...82F}. The extrapolated data are presented as 27-day moving averages, with the plotted error being a 99\% confidence interval. The magnetofrictional constant in the Outflow Fields is $\nu_0 = 5 \times 10 ^ { -17} \rm s \, cm ^{-2}$ and the source surface height is $r_{ss} = 2.5R_\odot$.}
\label{fig:5_plot_time}
\end{figure*}

Figure \ref{fig:5_plot_time} shows a comparison between these in-situ data and four series' of magnetic field extrapolations, calculated using lower boundary magnetic field data representing every day between 2000 and 2022 (the range over which both the MDI/HMI magnetograms are reliable and the data from \citet{2022SoPh..297...82F} exists). Similarly to the previous section, we vary the sound speed $c_s$ and fix the magnetofrictional constant $\nu_0$, although similar results could be obtained by doing the opposite (in which case increasing $\nu_0$ will generally increase the OSF). 

The lower boundary data are prepared using the process described in Section \ref{sec:heightvary}. We combine data from two instruments: the MDI instrument on SOHO \citep{1995SoPh..162..129S} and the HMI instrument on SDO \citep{2012SoPh..275..207S}. The MDI data are used for Carrington rotations before CR2098 (2010 June). The data from the two instruments are not naturally calibrated, and so we scale the HMI data to match MDI by the factors given in \citet{2012SoPh..279..295L}. It would of course be justifiably valid to instead take the HMI data as the ground truth and scale MDI as appropriate, but this would make the OSF discrepancy more extreme and so we follow \citet{2019SoPh..294...19W} by choosing to scale HMI. Despite HMI being online for more than 15 years, discussions of such calibration and data preparation are ongoing. Indeed, \cite{2025ApJS..278...55P} and \cite{2024A&A...690A.341S}  suggest that HMI magnetograms are consistent underestimates of the true magnetic field strength, with the former suggesting this discrepancy may even be up to a factor of two. The spatial resolution of the data also makes a large difference when modelling small-scale regions \citep[eg.][]{2024A&A...683A.134M}, but this has less of an effect on a global scale as the higher-order harmonics do not persist high into the corona.

We see, as expected, that the PFSS extrapolation underestimates the measured value of OSF -- on average over this range the predicted fluxes from PFSS are around $45\%$ lower than the expected values. The Outflow Fields clearly perform better, with the field calculated using $c_s = 150\,\mathrm{km}\,\mathrm{s}^{-1}$ shown here matching the target very closely.

\begin{figure*}
\plotone{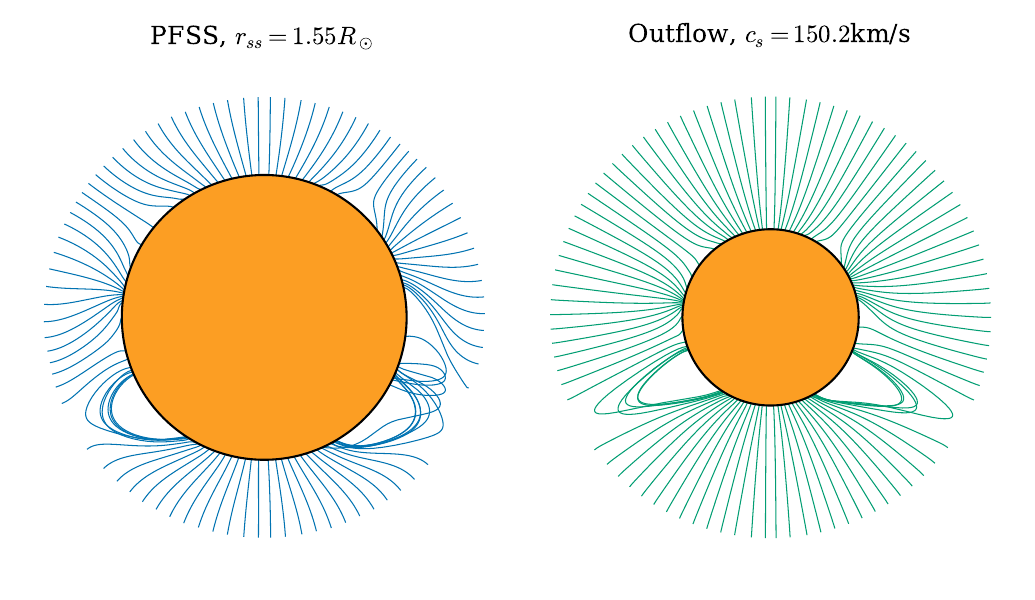}
\caption{Examples of matching OSF measurements with no regard for the topology of the magnetic field. The left panel shows the magnetic field lines for a PFSS field with a low source surface ($r_{ss} = 1.55R_\odot$) and the right panel shows an Outflow Field with $c_s = 150.2\,\mathrm{km}\,\mathrm{s}^{-1}$ and $r_{ss} = 2.5R_\odot$. Both of these fields use the HMI data from 2017 August 21 as a lower boundary, and match the measured in-situ OSF perfectly at that time.}
\label{fig:6_plot_sillylines}
\end{figure*}

A perhaps more useful measure of the accuracy of these extrapolations is the cross-correlation coefficient $r$ between the sets of data, which will be reliable despite any overall discrepancies in the scaling of the input data. For this data set the PFSS prediction correlates with the in-situ data with $r = 0.748$. All three sets of Outflow Fields shown here perform better than this, with the best-correlated series of fields being with the sound speed $c_s = 150\,\mathrm{km}\,\mathrm{s}^{-1}$, where $r = 0.864$. This indicates that irrespective of any overall discrepancies in the magnitude of the input magnetic field data, the Outflow Fields more accurately predict the rise and fall in OSF throughout the Solar Cycle, relative to a PFSS equivalent.

It is vital to note, however, that altering the magnetic field topology with only the goal of matching OSF values is prone to result in magnetic field structures which appear unrealistic relative to other observations. This applies both to the tried-and-tested approach of reducing source-surface heights in PFSS but also to increasing the outflow speed in our new model. This is shown to full effect in Figure \ref{fig:6_plot_sillylines}, which shows a PFSS and Outflow Field, each with  parameters chosen such that the OSF matches the target exactly for the date of the 2017 solar eclipse. Comparing to the eclipse photograph in Figure \ref{fig:7_real_images}(a), we see that both modelled field topologies are clearly very different from the white-light observation.

To combat this problem, the remainder of this paper is dedicated to optimising our Outflow Field parameters based on the topology of the coronal magnetic field observed during solar eclipses. The shape of these structures is independent of any overall bias in the magnetic field strength measured by HMI, MDI or other magnetograms and should provide a more objective comparison.

\section{Optimization of Outflow Field extrapolations with respect to field line topology}
\subsection{Identification and classification of magnetic field lines in eclipse photographs}

\begin{figure*}
\plotone{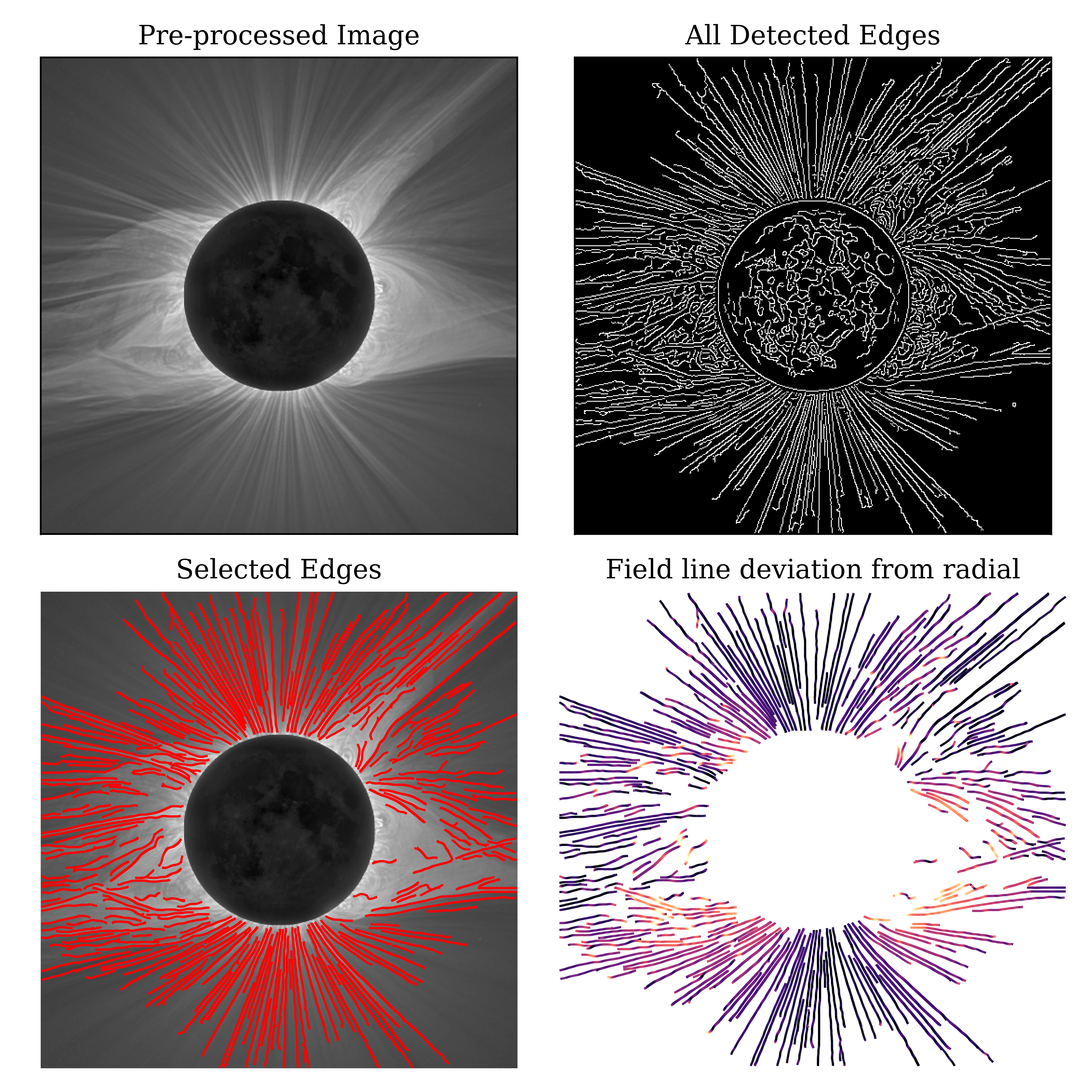}
\caption{The edge detection process for an image of the inner solar corona taken during the August 2017 Solar Eclipse. The original image is cropped to 2.5 solar radii in each direction, converted to greyscale and sharpened (top left), before applying the Canny edge detection algorithm (top right). These edges are then filtered based on their applicability (bottom left) -- they must be sufficiently long to reliably measure angles. The deviation between the angle of the traced field lines and the radial direction is shown the bottom right panel, where the colourmap ranges from purely radial (black) to around $60^{\circ}$ deviation (yellow/white).}
\label{fig:7_real_images}
\end{figure*}

In principle, comparing the topology of the corona is far more difficult than comparing a scalar value like the total OSF. The approach we have taken is to compare our Outflow Fields to a series of 12 photographs of Solar eclipses taken and processed by C. Emmanoulidis and M. Druckmüller \citep{2006CoSka..36..131D}, ranging from 2006 to 2024. From these photographs we aim to identify the shape of the magnetic field lines visible from the direction of Earth, specifically their angle relative to the radial direction. We can then compare against magnetic field lines traced numerically from the outputs of our model for a variety of outflow speeds, which can be altered to match the photographic `ground truth' as closely as possible for each eclipse. This then provides an objective best fit for the Outflow Field parameters completely independently of any measurements of the OSF, and of the associated uncertainties.

Figure \ref{fig:7_real_images} shows the process we use to identify the magnetic structures visible in the Solar Eclipse of August 2017. The same process was applied to 12 eclipse photographs in total -- all of the total eclipses between 2006 and 2024 except 2021 and 2023. We were inspired in part by the work of \citet{2020ApJ...895..123B} which showed that such analysis is possible. 

The top left panel of Figure \ref{fig:7_real_images} shows the first stage of processing for the 2017 eclipse image. We first convert each full colour image to greyscale, interpolate to a resolution of $512 \times 512$ and linearly scale the overall brightness to be consistent for all images. The brightness we found to be ideal for these purposes is a mean pixel value of $105$ (where the maximum is $256$). We also paste a copy of the same image of the Moon's face onto each photograph to provide a more objective comparison between each of them.

\begin{figure*}
\plotone{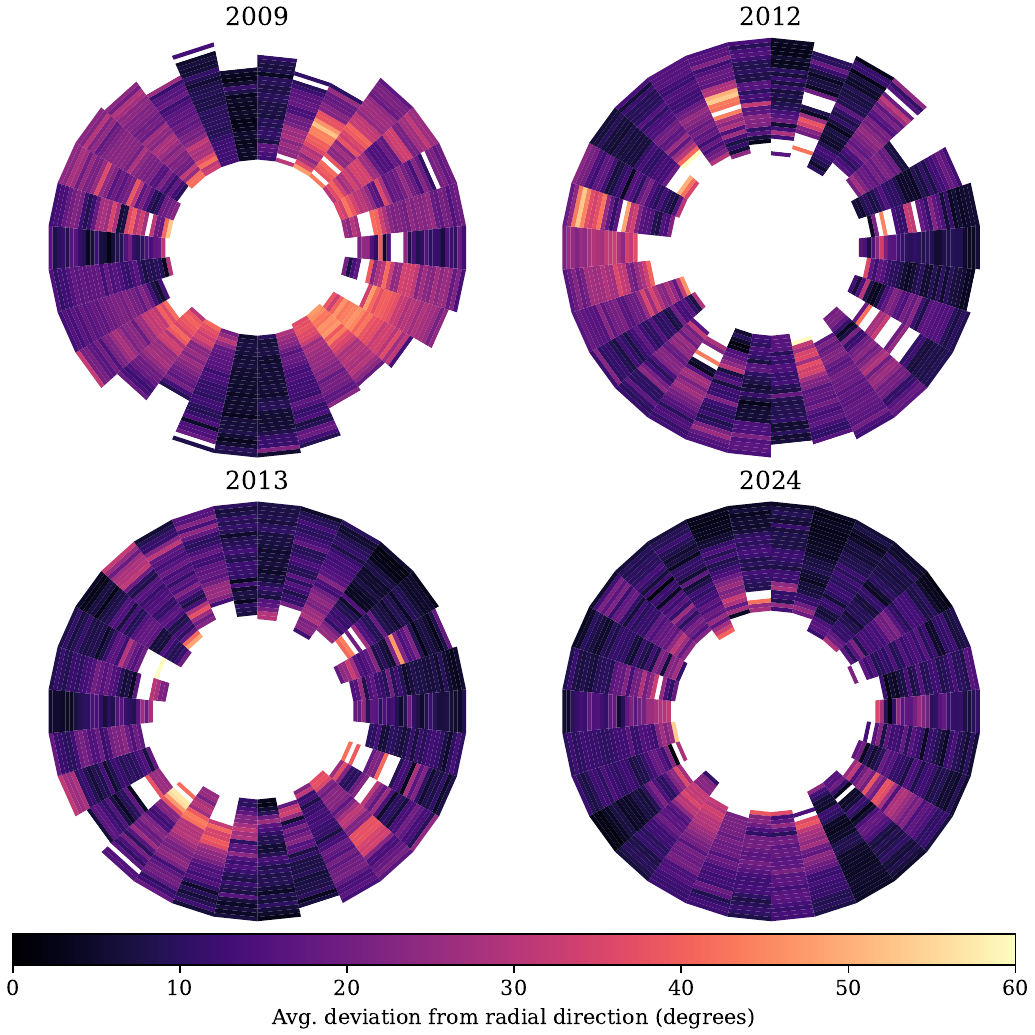}
\caption{Histograms of the average field line direction relative to the radial direction, calculated directly from four eclipse photographs. There are 30 angular and radial bins up to $2.5R_\odot$. The unfilled cells are those without sufficiently many identified field lines.}
\label{fig:8_histograms}
\end{figure*}

We then apply the Canny edge detection algorithm \citep{4767851}, implemented using the Python OpenCV package \cite{opencv_python}. This results in the image shown on the top-right of the figure, with edges identified in white. Clearly several of these are unrelated to the magnetic field structure, and several more are too short/noisy to be of any use for comparison, especially near the base of the large streamers on either side of the image. In their paper, \citet{2020ApJ...895..123B} do manage to analyse the magnetic field in these regions by using higher-resolution images for some of the eclipses, and combining them with the wider-field images for the overall statistical analysis.

We thus filter the identified field lines by their length and the locations of their endpoints. The chosen requirements are that a field line must be more than 20 pixels long and lie in part between $1.05$ and $2.45$ solar radii in altitude. The 283 lines which satisfy these criteria are shown in red on the lower-left panel of Figure \ref{fig:7_real_images}. 

It is undoubtedly possible to use a more sophisticated approach to identifying the overall topology of the field, but for our purposes we follow \citet{2020ApJ...895..123B} and \cite{2025arXiv250313292R} in merely identifying the field line angles with respect to the radial direction, and classifying them as such. To reduce noise, the field line `coordinates' are smoothed slightly (the finite $512\times 512$ pixel grid would result in far too many sharp corners otherwise) and the angle of the field line at each point is then calculated. These angles (or more accurately the angle relative to the radial direction) are plotted in the lower right panel of Figure \ref{fig:7_real_images}. At the time of the 2017 eclipse the field is relatively dipolar, and thus we see that near the poles the field is almost entirely radial (black), whereas near the equator and lower in the corona the field line angles deviate significantly, up to around $60^{\circ}$ on this plot (in yellow).

To classify this distribution objectively we further follow \citet{2020ApJ...895..123B} by splitting the region between $1-2.5 R_\odot$ into bins, and finding the average field line direction relative to radial for each bin. We choose $30$ equally spaced bins in both the radial and angular directions. The resulting distributions after this process are shown in Figure \ref{fig:8_histograms}, for the 2009, 2012, 2013 and 2024 eclipses. We can clearly see the relation between the distribution of the $2017$ eclipse and the field lines on the bottom right of Figure \ref{fig:7_real_images}. Note that the distribution of field line angles varies wildly between the eclipses -- this is primarily due to the fact we are observing the magnetic field at different stages in the solar cycle. In the period of the cycle where the Sun is most active (such as $2024$ here), the field lines are on average more aligned with the radial direction, and in general there is no clear dipolar structure, unlike when the Sun is quiet.

\begin{figure*}
\plotone{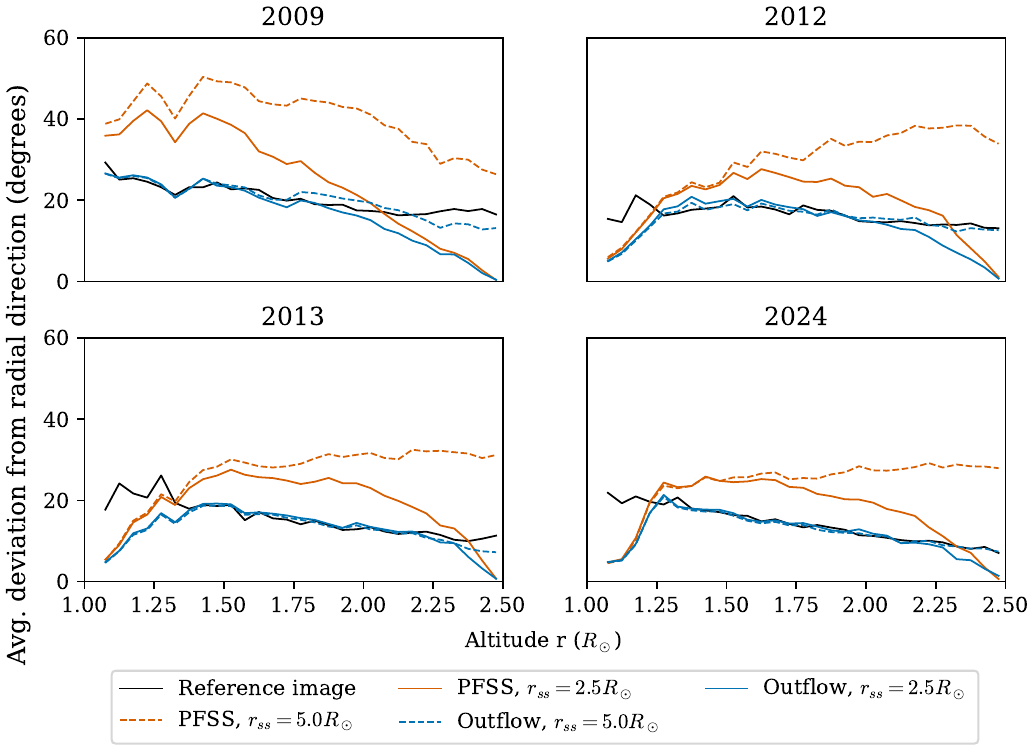}
\caption{Average field line angles relative to the radial direction. Showing (in black) the field lines detected on four eclipse images and the equivalent distributions obtained from PFSS and optimized Outflow Fields with source surface heights at $2.5R_\odot$ and $5.0R_\odot$. The 2009/2013 eclipses were the worst matches, and 2012/2024 were the best.}
\label{fig:11_quality_check}
\end{figure*}

\begin{center}
\begin{table*}[ht!]

\begin{tabular}{ccccccc}
\hline
\multicolumn{1}{c}{} & \multicolumn{2}{c}{PFSS} & \multicolumn{2}{c}{Outflow: Optimized Topology} & \multicolumn{2}{c}{Outflow: $c_s = 150 \mathrm{km}\,\mathrm{s}^{-1}$} \\
 \hline
 Eclipse Year & $r_{ss} = 2.5R_\odot$ & $r_{ss} = 5.0R_\odot$ & $r_{ss} = 2.5R_\odot$ & $r_{ss} = 5.0R_\odot$ & $r_{ss} = 2.5R_\odot$ & $r_{ss} = 5.0R_\odot$ \\
 \hline 
2006 & 7.078 & 16.930 & 5.558 & 1.994 & 10.890 & 10.866\\
2008 & 7.924 & 20.124 & 5.447 & 1.094 & 11.916 & 12.118\\
2009 & 11.106 & 21.523 & 6.433 & 2.184 & 12.556 & 12.705\\
2010 & 9.538 & 23.185 & 3.891 & 1.119 & 7.940 & 7.989\\
2012 & 6.813 & 17.230 & 4.131 & 0.887 & 7.402 & 7.501\\
2013 & 8.177 & 15.889 & 3.460 & 2.518 & 6.884 & 6.925\\
2015 & 4.738 & 15.714 & 4.631 & 1.001 & 9.144 & 9.254\\
2016 & 3.992 & 15.626 & 3.584 & 2.089 & 8.088 & 8.012\\
2017 & 5.694 & 18.349 & 4.340 & 1.192 & 11.243 & 11.371\\
2019 & 6.464 & 18.881 & 5.537 & 0.995 & 13.686 & 13.735\\
2023 & 6.623 & 20.041 & 3.191 & 1.318 & 6.165 & 5.963\\
2024 & 7.271 & 14.508 & 1.878 & 0.815 & 5.288 & 5.234\\
  \hline
\end{tabular}

\caption{Table showing the standard deviation of the error between the average field line angles of the various models, and the eclipse image `ground truth', measured in degrees and for altitudes above $r=1.25R_\odot$. This quantifies the deviations shown in  Figure \ref{fig:11_quality_check}. We also include the equivalent metric for Outflow Fields using a Parker wind profile with $c_s = 150 \mathrm{km}\,\mathrm{s}^{-1}$.}

\label{table:deviationtable}
\end{table*}
\end{center}

As a final step we take the mean values of each of the bins at a given altitude, to give an angle distribution in altitude shown as the black lines in Figure \ref{fig:11_quality_check} (for four of the eclipses), disregarding those bins in which too few field lines were identified. We see that at this time the average field line angle relative to the radial direction is around $20^\circ$ 
near the solar surface, falling to around $10^\circ$ at $2.5R_\odot$ (a purely radial field would measure as $0^\circ$). This is the general pattern for all the eclipses, albeit with some variation. Without direct comparison to models (as we shall do imminently), the most notable feature visible here is that the field lines are definitely not radial at the commonly-used source surface height of $2.5R_\odot$.

\subsection{Comparisons with PFSS and Outflow models}

\label{sec:comps}

To provide an objective comparison of the field line topologies between the eclipse photographs and our models, we use boundary data gathered at the precise time of each eclipse, processed as described in Section \ref{sec:heightvary}. Our code allows for the calculation of both PFSS and Outflow Fields, with arbitrary source surface heights and outflow speeds (PFSS is just an Outflow Field with $V(r) = 0$), and has the built-in option to trace and save field line coordinates with arbitrary start points.

We have freedom here to select the distribution of these start points. We do so based on the principle of the `Thomson Surface' \citep{2012ApJ...752..130H}, which (to a first approximation) predicts the relative brightness of visible emissions from the magnetic field, as a function of the angle of the magnetic field relative to the observer. As shown in Figure 1 of \citet{2012ApJ...752..130H}, areas of the magnetic field which are close to forming a right angle between the Sun and the Earth are most bright, with the brightness roughly following a $\sin^2(\theta)$ distribution, where $\theta$ is this angle. Thus we pick the field line seeds randomly in the radial and angular (in the plane facing the observer) direction, and in the third dimension weight them based on the Thomson scattering angle. 

The results of this process are shown in orange on Figure \ref{fig:11_quality_check}, showing two PFSS fields for each of four eclipses, using either the standard source surface height of $r_{ss} = 2.5R_\odot$ or increased to $r_{ss} = 5.0R_\odot$. We see that for the former, the field lines become radial at $r_{ss} = 2.5R_\odot$ as expected -- this is precisely the upper boundary condition. For the higher source surface, however, the magnetic field structures are vastly different, and the average deviations from the radial direction are far higher than the reference photographs. It is also clear that generally no other PFSS source surface height would result in a distribution more closely matching the reference images (black). This exemplifies the need to introduce the solar wind outflow to modify the field structure, rather than merely changing $r_{ss}$. Similar patterns were seen in the other eight eclipse photographs we considered.

\subsection{Optimization of the outflow model parameters based on observations} 
\label{sec:optimisation}

Naturally the average deviation from radial for each field line is insufficient to capture the full complexity of the magnetic field structure, but we have shown that as a measure it is capable of identifying some of the major flaws with the PFSS model. We can also use this measure to optimize our outflow model to match the field line angle distributions as closely to observations as possible. We essentially have the freedom to choose any outflow profile we desire, and are no longer limited to the Parker profiles such as those in Figure \ref{fig:1_outflow_speeds}.
To determine the `ideal' outflow profile for each eclipse, we use a Covariance Matrix Adaptation Evolution Strategy (CMA-ES) \citep{CMA-ES}, ideally suited to problems such as these where small-scale changes to the input parameters can have unpredictable and surprisingly large effects. We wish to constrain the profile to be non-negative and monotonically increasing, and so define $V(r)$ as
\begin{equation}
    V(r) = \frac{p(r)e^{p(r)}}{e^{p(r)} + 1}
\end{equation}
where $p(r)$ is an arbitrary 4\textsuperscript{th}-order polynomial. These solutions are additionally numerically constrained such that they attain a constant value equal to the maximum speed, if they would otherwise be decreasing with altitude.

We set up an optimization run for each of the 12 eclipses individually, starting with a reasonable guess for the polynomial $p(r) = -5r + 4$. We use the `cma' python package for the optimization \citep{cma}, which seeks to minimize the squared error between the radial distributions of the magnetic field lines relative to the reference for altitudes above $1.25R_\odot$ (comparing the colored lines on Figure \ref{fig:11_quality_check} with the black, essentially). 

The optimized values of this measure are given in Table \ref{table:deviationtable}, for all $12$ of the considered eclipses. Comparing with the equivalent measure for PFSS fields, we note that Outflow Fields optimized for topology outperform PFSS in every case, for both source surface heights of $r_{ss}=2.5R_\odot$ and $r_{ss}=5.0R_\odot$. In the table we also present the equivalent metric for Outflow fields using a Parker profile with  $c_s = 150 \mathrm{km}\,\mathrm{s}^{-1}$, which on average matches the measured OSF well (as seen in Figure \ref{fig:5_plot_time}). We see, as expected, that the topology of these fields do not correspond well with the eclipse observations, often performing worse even than PFSS. We find that for all the eclipses the algorithm converges to a solution in around 1000 iterations, roughly a few hours' computation time on a single core. The outflow calculations are on a $360 \times 180 \times 120$ grid, calculated up to $r_{ss} = 5.0R_\odot$, to allow for the clearly non-radial field line distribution at $r_{ss} = 2.5R_\odot$ in the observations.

\begin{figure*}
\plotone{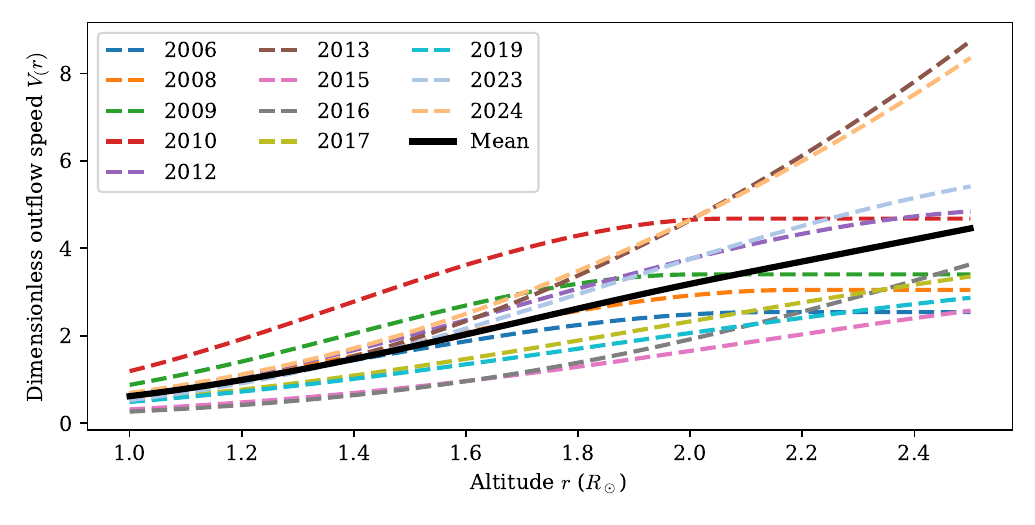}
\caption{Optimum profiles for the outflow speeds, for each of the $12$ eclipses we consider. These speeds are chosen to match the field line angular distribution with each of the solar eclipse photographs in turn. The mean profile over all the eclipses is shown as the thick black line.}
\label{fig:10_optimums}
\end{figure*}

The resulting outflow speed profiles after this optimization process are shown in in Figure \ref{fig:10_optimums}, for each of the 12 eclipses we consider. All are roughly the same order of magnitude, which is heartening given there were no restrictions on the parameter space. We note that most eclipses (except 2009 and 2010) have low outflow speeds close to the solar surface, rising to between 2.5 and 9 code units by $r_{ss} = 2.5R_\odot$. Note that we deliberately express these speeds in terms of the nondimensional $V(r)$ rather than physical units, as the profiles are determined entirely empirically without needing to assume a value for the magnetofrictional constant. For reference, if we do assume a constant of $\nu_0 = 5 \times 10^{-17} \rm s \, cm^{-2}$, $5$ code units (roughly the average value at $2.5R_\odot$) corresponds to an outflow speed of $14\,\mathrm{km}\,\mathrm{s}^{-1}$. It must be stressed that this value can be changed arbitrarily by altering the magnetofrictional constant, however, and as such only the `code units' $V(r)$ is meaningful. 

A more realistic value of the solar wind speed at this altitude would likely be closer to $100 \,\mathrm{km}\,\mathrm{s}^{-1}$, which would require reducing $\nu_0$ to  around $7.2 \times 10 ^ {-18}\,\rm s \, cm^{-2}$. This substantial reduction may have consequences for time-dependent magnetofrictional modelling, although this remains to be explored.

The fields from $2013$ and $2024$ have optimum speeds which increase throughout the domain (and indeed beyond $2.5R_\odot$), whereas other eclipses, such as $2010$ and $2009$, are relatively fast low in the domain but level off at around $2.0R_\odot$, above which the wind speed is constrained to be constant (as the raw polynomial $p(r)$ will decrease, and this is nonphysical).

We note that the fastest overall wind speeds are for the 2013, 2024 and 2023 eclipses, all of which occurred at times close to Solar Maximum, and the lowest were at times closer to Minimum, but despite this there is very little correlation overall. The correlation between the true solar wind speed and the solar cycle is not entirely clear, in part because the real speed varies in latitude and longitude, and across spatial scales. \citep[e.g.][]{2009LRSP....6....3C, 2012ApJ...751..128M, 2021A&A...654A.111D}. In any case, we are not claiming it is possible to determine the correct solar wind speeds directly from our model (as magnetofriction does not take into account nearly enough of the physics), but merely which of the input `speeds' match observations most closely. A number of other factors may likely have an effect on which of these eclipses require higher solar wind speeds, perhaps most obviously the large variation in streamer sizes between different eclipses.

The field line angles for the optimum solutions for four eclipses are shown on Figure \ref{fig:11_quality_check} in blue, for both $r_{ss} = 5.0R_\odot$ (used in the optimization run) and the more commonly-seen $r_{ss} = 2.5R_\odot$. We observe that for the former the average field line angles match the reference very well above $1.4R_\odot$. Below this height the lack of reliable field lines identified in the reference image is likely the issue with the remaining discrepancy, which we note is still much less than the equivalent with PFSS fields. The Outflow Fields with $r_{ss} = 2.5R_\odot$ match the reference nicely for mid-level altitudes, but being constrained to be radial at the top boundary naturally makes this solution less realistic near the top of the domain.

\subsection{OSF predictions from optimized Outflow Fields}

For completeness, it remains to judge the performance of the OSF predicted by these optimized Outflow Fields, relative to the in-situ measurements as in  Section \ref{sec:frostmatch}. As stated then, to match the OSF measurements exactly requires quite a fast outflow speed, which in itself leads to unrealistically radial magnetic field lines. This can be seen when comparing the right panel of Figure \ref{fig:6_plot_sillylines} with the eclipse photograph in Figure \ref{fig:7_real_images}, in which we see that the field lines are very rarely purely radial (apart from at the poles, in this particular case). Thus the OSF predictions from the optimized model are still in general underestimates. 

\begin{figure*}
\plotone{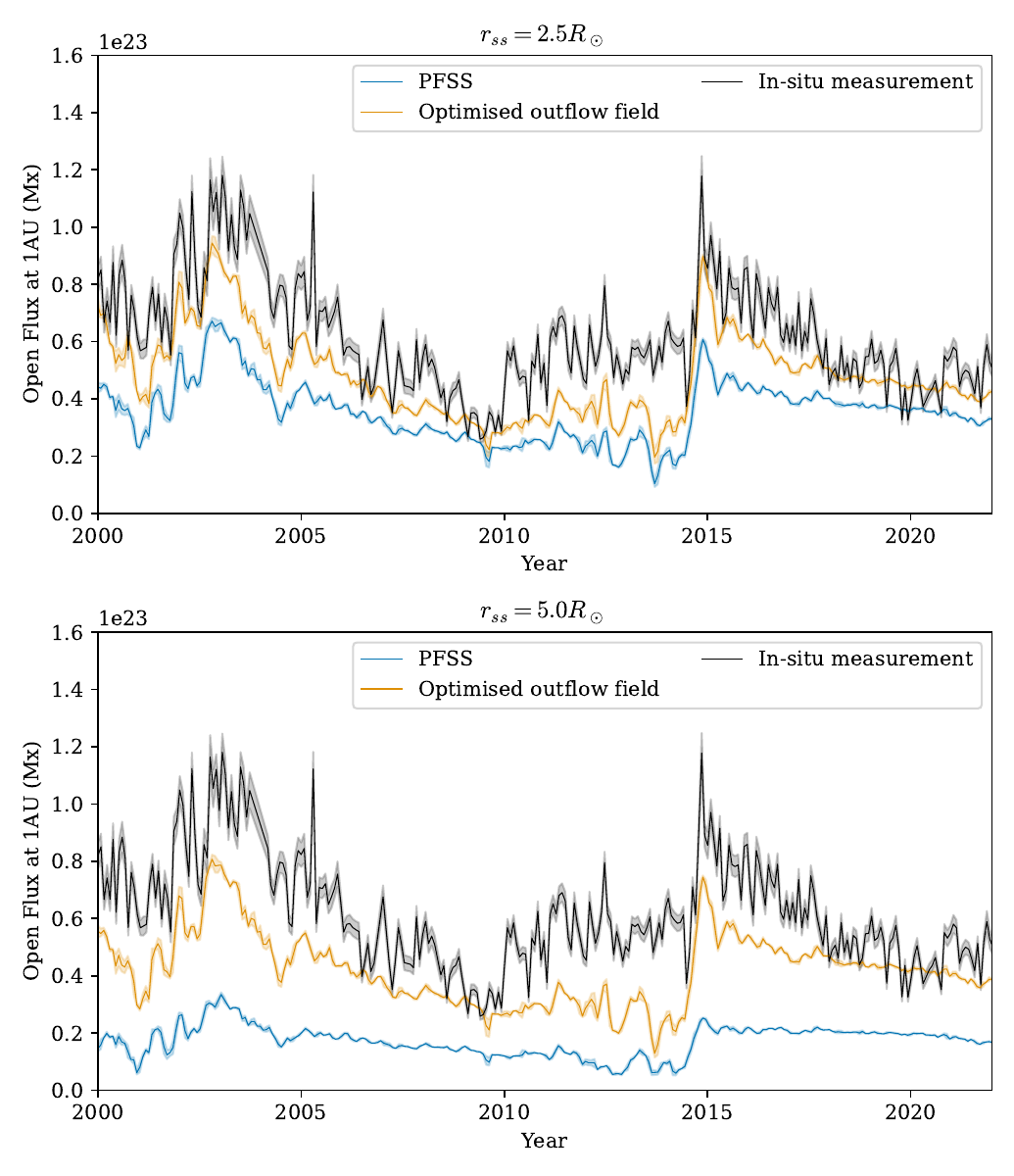}
\caption{Comparison between PFSS, Outflow Field extrapolations and in-situ measurements of the OSF at 1AU. The in-situ data are taken from \citet{2022SoPh..297...82F}. The extrapolated data are presented as 27-day moving averages, with the error range being a 99\% confidence interval. The Outflow Fields use the mean optimum solar wind speed for all 12 eclipses. The source surface height is $r_{ss} = 2.5R_\odot$ on the upper panel, and $r_{ss} = 5.0R_\odot$ on the lower.}
\label{fig:13_plot_optimums}
\end{figure*}

This is shown in Figure \ref{fig:13_plot_optimums}, which (similarly to Figure \ref{fig:5_plot_time}) compares the ground-truth in-situ OSF measurements from \citet{2022SoPh..297...82F} against PFSS and Outflow models. For a fair comparison we plot the data for fields with both $r_{ss} = 2.5R_\odot$ and $r_{ss} = 5.0R_\odot$.

The difference here is that the Outflow model uses an outflow speed profile which is the mean of all the individually-optimized eclipse observations (the black line on Figure \ref{fig:10_optimums}). We use this as the default profile for Outflow Fields in the `outflowpy' package. As is well known and expected \citep{2017ApJ...848...70L, 2019SoPh..294...19W}, the PFSS fields underestimate the target value significantly.

 Using the standard source surface radius of $r_{ss} = 2.5R_\odot$, for the PFSS fields the shortfall is $45\%$ (over this particular time range), whereas the optimized outflow model still underestimates the true value, but only by $24 \%$. When increasing the source surface height to $r_{ss} = 5.0R_\odot$ the PFSS fields perform far worse, underestimating the target by $74\%$. Raising the source surface height has less of an effect on the outflow fields, where the underestimate is a comparatively low $34 \%$. Hence the use of outflow fields (relative to PFSS) goes some way, though definitely not all the way, to alleviating the so-called Open Flux Problem \citep{2017ApJ...848...70L, 2019SoPh..294...19W}.

\section{Conclusions}

\begin{figure*}
\plotone{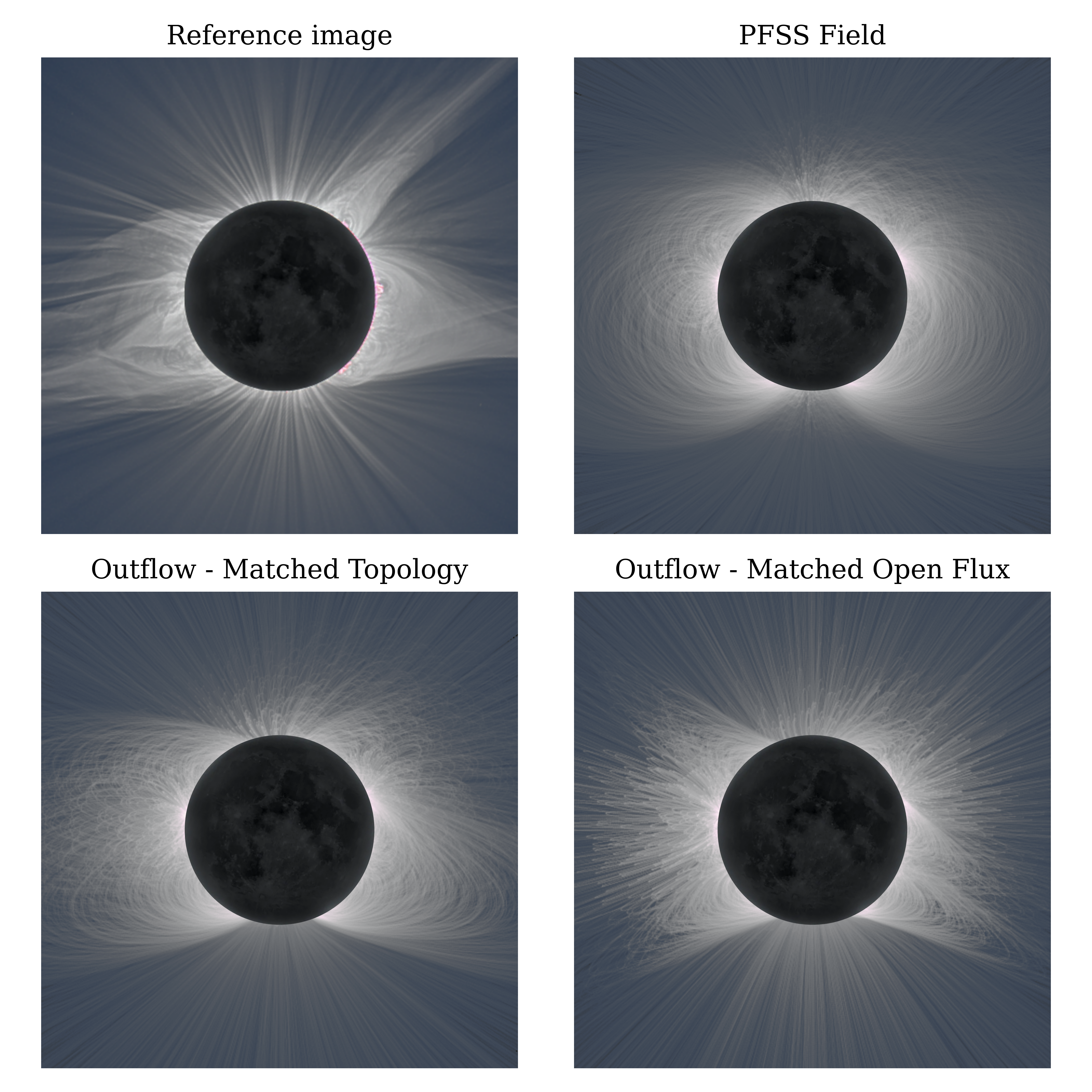}
\caption{Qualitative comparison between PFSS, two Outflow Fields, and an image of the 2017 Solar Eclipse (courtesy of C. Emmanoulidis and M. Druckmüller). The synthetic images of the models are generated using the process described in Appendix \ref{sec:appendix}. All three use a source surface height of $r_{ss} = 5R_\odot$. The bottom-left panel shows an Outflow Field with the mean optimum wind speed profile obtained as described in Section \ref{sec:optimisation}, whereas the bottom-right panel shows an Outflow Field with parameters chosen to match the in-situ OSF measurements exactly.}
\label{fig:12_nice_image}
\end{figure*}

Following from our paper introducing Outflow Fields (Paper I), we have discussed more comprehensively the advantages these fields have over the well-known and used PFSS model. Outflow Fields self-consistently take into account the effect of the Solar Wind, which generally increases the open solar flux (OSF) relative to potential field equivalents by stretching magnetic streamers outwards and causing the magnetic field to become more aligned with the radial direction, especially at high altitudes. 

Improving upon our original paper, we have shown how one can generate a basic solar wind outflow profile expressed in terms of two free variables -- the solar wind speed $c_s$, which determines the `shape' of the solution and the magnetofrictional constant $\nu_0$, which essentially acts as an overall scaling factor. 

We believe that there are several advantages of using these fields in favor of PFSS models. One of the most significant is that the overall OSF predicted by Outflow models is roughly independent of the (somewhat arbitrary) source surface height $r_{ss}$. Whilst there have been several works seeking to optimize PFSS models by changing this height, this often leads to unrealistically-shaped magnetic field structures. In Outflow models, changing $r_{ss}$ has little effect for reasonable parameters, and so we can instead vary the outflow speed to achieve this instead.

Secondly, the opening-out of the field lines increases the OSF such that if desired, the solar wind speed can be increased to a point at which the predicted OSF values match those measured in-situ exactly. Alas, this high speed leads to discrepancies between the magnetic field structure in these solutions and equivalent observations of the magnetic field during eclipses. However, the the established alternative approach of reducing $r_{ss}$ to below $2.0R_\odot$ performs far worse.

Finally, and most significantly, we have introduced a method which optimizes the solar wind profile to match the observed structure of the magnetic field during eclipses as closely as possible. We do this by comparing the angles of observed magnetic field lines against the equivalent in our extrapolations based on magnetogram data at the same times. We have shown that PFSS models cannot match the structure well for any $r_{ss}$ (perhaps most obviously as the real magnetic field does not become radial at any particular fixed altitude). However, when optimal parameters are found our Outflow model does match this angular distribution quite closely at all altitudes above around $1.4R_\odot$ (below which the magnetic field lines in photographs are difficult to make out).

To illustrate this qualitatively, in Figure \ref{fig:12_nice_image} we present synthetic images of PFSS and Outflow Fields, along with a reference photograph from the 2017 eclipse. These images are generated using the process described in Appendix \ref{sec:appendix}. All of these images are generated from fields with source surfaces set at $r_{ss} = 5.0R_\odot$, and we see that the PFSS field, as usual, results in streamers which extend all the way up to this altitude, or in this case at least as high as the image shows. In the Outflow models, the streamers extend to between $1-3 R_\odot$, which is much more similar to those seen in the photograph. We also see here that when matching the OSF precisely the open field lines are unrealistically radial. Thus when choosing between whether to match field line topology or match OSF, we favor the former, which is the default in the `outflowpy' package. 

 One area in which the Outflow Fields underperform relative to PFSS is the shape of the magnetic field near the poles, which fan out more significantly in the images than the model predicts. This may possibly be alleviated in future by allowing the solar wind speed to depend on latitude, but that is currently beyond the scope of the model, which assumes an outflow speed dependent only on altitude. Such latitudinal dependence can be investigated by allowing a time-dependent magnetofrictional simulation to relax to an equilibirum state, but this cannot be done as quickly as an Outflow or PFSS field.

    Ultimately, when using a solar wind speed profile chosen to match field line shapes and with a source-surface height of $2.5R_\odot$, the outflow model still underestimates the OSF relative to in-situ measurements by around $24\%$ when calculated daily between $2000$ and $2022$. This is not ideal, but is a considerable improvement over PFSS, for which the equivalent discrepancy is $45\%$. As the scaling of the solar surface data from HMI/MDI is somewhat arbitrarily determined (given the two instruments differ by a scale factor of $1.4$ in any case, and we have arbitrarily chosen MDI as the ground truth to match most existing literature), we assert that the fields generated by matching the field line structure are likely the more accurate. 

This indicates that there are likely additional sources of OSF which we have not considered, such as the effects of non-potential structures in the lower corona \citep[flux ropes, CMEs etc.;][]{2006JGRA..11110104O} or from dynamic activity near open/closed flux boundaries \citep{2024ApJ...964..115A} resulting in magnetic fields emanating from these regions (which would otherwise be neglected) counting towards the OSF. Combining this with the Outflow model would almost certainly reduce this difference even further. Another approach to reducing the OSF discrepancy is to consider helicity condensation, as in \cite{2025ApJ...987...32K}, who deduce that around half the OSF shortfall relative to PFSS can be accounted for by dynamic processes in the lower corona. 

Alternatively, there may be some grounds behind considering the calibration of the instruments used for the lower boundary data \citep[e.g.][]{2014SoPh..289..769R}. As stated earlier, measurements from the HMI and MDI instruments differ by a factor of $1.4$ -- applying such a correction once again \citep[which may be not unreasonable, as suggested by][]{2025ApJS..278...55P} could solve the Open Flux problem. 

In conclusion, we have objectively established optimal parameters for the Outflow model, which can be used as an alternative to the established PFSS fields as a first-order approximation the global coronal magnetic field. To this end, and for ease of use of our model, we have released a python package `outflowpy', which is designed to be easily compatible with the existing `pfsspy' package by David Stansby \citep{david_stansby_2023_8280209}. We thus hope that these fields can be of use to the solar community in a wide variety of applications.

\begin{acknowledgments}
We thank Prof. Miloslav Druckmüller of Brno University for the use of his solar eclipse photographs, and Mathew Owens for his help with obtaining the strahl-based OSF estimates from \citet{2022SoPh..297...82F}. We also thank Yihua Li and collaborators for identifying the inaccuracy of the Solar Wind speed approximation in \citet{2021ApJ...923...57R} at high solar wind speeds. This work was supported by the Science and Technology Facilities Council [grant number UKRI1216].
\end{acknowledgments}

\begin{contribution}

OEKR wrote the code, performed the calculations and created the package `outflowpy' under the close guidance of ARY, who conceived the project and determined the research plan. Both authors contributed to writing the paper.

\end{contribution}

%



\appendix

\section{Generation of Synthetic Eclipse Images}
\label{sec:appendix}
We here describe the process behind generating the synthetic eclipse images in Figure \ref{fig:12_nice_image}. Such white-light synthetic images are commonplace and relatively easy to produce from models such as MHD, when the density and temperature of the corona are known. They can also be produced from models which lack these features, such as dynamic MF models of active regions, by regarding the electric current density. Alas for PFSS/Outflow fields none of these approaches are suitable.

Instead, we have developed a new approach, based on tracing a large number of field lines and weighting their `emissions' by factors determined by comparison with reference eclipse images. To generate the images in Figure \ref{fig:12_nice_image}, we trace $25,000$ field lines. The field lines are traced in both directions from start seeds sampled randomly using a Latin Hypercube distribution, with equal weightings in $\cos \theta$  (latitude) and longitude $\phi$. The sampled altitude of the start seeds is skewed such that there are more start points lower in the domain.

At each point which each field line exists, an `emission' is added to the overall image matrix. The magnitude of this depends on the Thomson Scattering angle $\alpha$, (see Section \ref{sec:comps}), the maximum height of the field line (relative to the top of the domain) $\beta$, the magnetic field strength at that point $\gamma$ and at the field line footpoints $\lambda$, both relative to the maximum magnetic field strength overall. These are combined as follows:

\begin{equation}
    E = \sin^2(\alpha) \times \beta^{-2.986} \times \gamma^{0.195} \times \lambda^{0.279}.
\end{equation}

Adding all these emissions together creates a raw greyscale image. The pixels of this are then scaled and colourised to match to a reference image (the 2017 eclipse image in the Figure) such that the distributions of pixel brightness are the same. The parameters in the above equation have been determined using the CMA-ES evolutionary algorithm, by comparing the synthetic images against the references for all 12 eclipses we consider. The `distance metric' used in the image comparison is a simple Mean Squared Error (MSE), after we determined that a machine-learning similarity approach was too prone to settling into local minima (although those images did pick out the polar fields more clearly).


\bibliography{sample}{}
\bibliographystyle{aasjournalv7}



\end{document}